\documentclass[a4paper,12pt]{article}

\usepackage{epsfig,graphicx,amsmath,amssymb}
\usepackage[sort&compress,numbers]{natbib}
\usepackage{url}
\usepackage[percent]{overpic}
\usepackage{siunitx}
\usepackage{units}
\usepackage{hyperref}
\usepackage{wrapfig}
\usepackage{epsfig}
\usepackage{graphicx}
\usepackage{subfig}
\usepackage{hyperref}
\usepackage{wrapfig}

\begin{document}

\thispagestyle{empty}

$\phantom{.}$

\vspace{-1.5cm}
\begin{flushright}
{\sf LTH 1294, MPP-2022-8 \\
  } 
\end{flushright}

\hfill

\vspace{-1.0cm}
\begin{center}
{\Large {\bf {Mini-Proceedings of the\\[1mm] STRONG2020 Virtual Workshop on\\[1mm] ``Space-like and Time-like determination of the\\[1mm]
Hadronic Leading Order contribution\\[1mm] to the Muon $\boldsymbol{g-2}$'' }}
\vspace{0.75cm}}

\vspace{1cm}

{\Large  Wednesday, 24 November 2021 -- Friday, 26 November 2021}

\vspace{1.cm}

{\it Editors}\\
Andrzej Kup{\'s}{\'c} (Uppsala), Graziano Venanzoni (Pisa)
\vspace{1cm}

\begin{figure}[h]
    \centering
    \includegraphics[width=5cm]{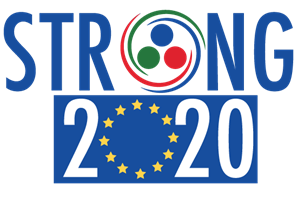}
\end{figure}

\vspace{1cm}

ABSTRACT

\end{center}

\vspace{0.3cm}

\noindent
The mini-proceedings of the STRONG2020 Virtual Workshop ``Space-like and Time-like determination of the Hadronic Leading Order contribution to the Muon $g-2$'', November 24--26 2021, are presented. This is the first workshop of the STRONG2020 WP21: JRA3-PrecisionSM: Precision Tests of the Standard Model (\url{http://www.strong-2020.eu/joint-research-activity/jra3-precisionsm.html}). The  workshop was devoted to review of  the working group activitity on: $(\it i)$ Radiative Corrections and Monte Carlo tools for low-energy hadronic cross sections in $e^+ e^-$ collisions; ($\it ii$) Annotated  database for $e^+e^-$ into hadrons processes at low energy; ({\it iii}) Radiative Corrections and Monte Carlo tools for $\mu$--$e$ elastic scattering.

\medskip\noindent
The web page of the conference:
\begin{center}
\url{https://agenda.infn.it/event/28089/} 
\end{center}
\noindent
contains the presentations.

\noindent

\newpage

{$\phantom{=}$}

\vspace{0.5cm}

\tableofcontents

\newpage

\section{Introduction}
\addtocontents{toc}{\hspace{1cm}{\sl A.~Kup{\'s}{\'c}, G.~Venanzoni}\par}

\vspace{4mm}

\noindent
 A.~Kup{\'s}{\'c}$^1$ and G.~Venanzoni$^2$

\vspace{4mm}

\noindent
$^1$Department of Physics and Astronomy Uppsala University, Sweden\\
$^2$INFN, Sezione di Pisa, Pisa, Italy\\

\vspace{4mm}

The importance of continuous and close collaboration between the experimental and theoretical groups is crucial in the quest for precision in hadronic physics. This is the reason why the Working Group on “Radiative Corrections and Monte Carlo Generators for Low Energies” (Radio MonteCarLow, see \url{http://www.lnf.infn.it/wg/sighad/}) was formed a few years ago bringing together experts (theorists and experimentalists) working in the field of low-energy $e^+e^-$ physics and partly also the $\tau$-lepton community. Its main motivation was to understand the status and the precision of the Monte Carlo generators (MC) used to analyze the hadronic cross section measurements obtained in energy scan experiments as well as  with radiative return method, to determine luminosities. Whenever possible specially prepared comparisons, i.e., comparisons of MC generators with a common set of input parameters and experimental cuts, were performed within the project. The main conclusions of this major effort were summarized in a report published in 2010~\cite{WorkingGrouponRadiativeCorrections:2010bjp}. During the years the WG structure has been enriched of more research topics and more groups joined.
The working group had been operating for more than 15 years without a formal basis and a dedicated funding. Recently parts of the program have been  included as a Joint Research Initiative (JRA3-PrecisionSM) in the group application of the European hadron physics community, STRONG2020 (\url{http://www.strong-2020.eu}, to the European Union, with a more specific goal of creating an annotated database for the low-energy hadronic cross section data in $e^+e^- $ collisions. The database will contain information about the reliability of the data sets, their systematic errors and the treatment of Radiative Corrections.
All these efforts have been recently revitalized by the new high-precision measurement of the
anomalous magnetic moment of the muon at Fermilab~\cite{Abi:2021gix,Albahri:2021ixb,Albahri:2021kmg,Albahri:2021mtf}, which, when combined with the final result from the Brookhaven experiment~\cite{bennett:2006fi}, shows a $4.2\sigma$ discrepancy with respect to the state-of-the-art theoretical prediction from the Standard Model~\cite{Aoyama:2020ynm} (mainly based on Refs.~\cite{Aoyama:2012wk,Aoyama:2019ryr,Czarnecki:2002nt,Gnendiger:2013pva,Davier:2017zfy,Keshavarzi:2018mgv,Colangelo:2018mtw,Hoferichter:2019mqg,Davier:2019can,Keshavarzi:2019abf,Kurz:2014wya,Melnikov:2003xd,Masjuan:2017tvw,Colangelo:2017fiz,Hoferichter:2018kwz,Gerardin:2019vio,Bijnens:2019ghy,Colangelo:2019uex,Blum:2019ugy,Colangelo:2014qya}; see Refs.~\cite{Miller:2007kk,Jegerlehner:2009ry,Jegerlehner:2017gek} for earlier reviews), including an evaluation of the leading-order hadronic-vacuum-polarization contribution from $e^+e^-\to\text{hadrons}$ cross-section data. Moreover, the recent high-precision lattice evaluation by the BMW collaboration~\cite{Borsanyi:2020mff} shows tension with the time-like data-driven determinations of $a_\mu^\text{HVP, LO}$, being $2.1\sigma$ higher than the Muon $g-2$ Theory Initiative data-driven value. This reinforces the need for new data (both time-like and space-like) as well as independent lattice calculations at a similar level of precision, especially in view of the tensions that would arise if indeed the data-driven determination needed to be revised substantially~\cite{Crivellin:2020zul,Keshavarzi:2020bfy,Malaescu:2020zuc,Colangelo:2020lcg}. 
During the workshop the recent updates on the following activities have been reviewed: ($\it i$)  Radiative Corrections and the Monte Carlo tools for low-energy hadronic cross sections in $e^+ e^-$ collisions; ($\it ii$) Annotated  database for the $e^+e^-$ into hadrons processes at low energy; ({\it iii}) Radiative Corrections and the Monte Carlo tools $\mu$--$e$ elastic scattering; towards the ambitious goal of achieving a full NNLO MC generator for time-like and space-like processes.

\newpage

\section{$\boldsymbol{R}$ measurements}

\subsection{Exclusive Measurement of Hadronic Channels between 2.0 and 3.08 GeV from BESIII}
\addtocontents{toc}{\hspace{2cm}{\sl X.~Zhou}\par}

\vspace{5mm} 

Xiaorong Zhou

\vspace{5mm}

\noindent
University of Science and Technology of China, Hefei 230026, P. R. China\\
State Key Laboratory of Particle Detection and Electronics, Hefei, 230026, P. R. China
 \\
\vspace{5mm}

The Beijing Spectrometers~(BESIII) experiment operating at the Beijing Electron-- Positron Collider~(BEPCII) has unique advantage to advance our knowledge of strong interaction at low energy. It has collected  a total integrated luminosity of $\sim30$~fb$^{-1}$ so far, mostly at the charmonium production and XYZ fine scan region. Among them, a unique data set with an integrated luminosity of $>650$~pb$^{-1}$ in the c.m.~energy
$\sqrt{s}$ from 2.0 to 3.08 GeV, the so-called continuum region with R value constantly around 2.25, can be used to study the baryon form factors and search for vector mesons. 

The electromagnetic form factors~(EMFFs) are fundamental properties of the baryons that connected to their charge and magnetization distributions. 
In time-like, the EMFFs can be studied via electron--positron annihilation, $e^{+}e^{-}\to B\bar{B}$, where $B$ denotes a spin-1/2 baryon. 
The cross section of such reaction can be expressed in terms of the electric and magnetic FFs, $|G_{E}|$ and $|G_{M}|$, respectively.
At BESIII, the EMFFs and the pair production cross sections of SU(3) octet baryons have been studied.
The proton EMFFs are studied with both ISR~\cite{BESIII:2019tgo,BESIII:2021rqk} and energy scan~\cite{BESIII:2015axk,BESIII:2019hdp} methods, with the EMFF ratio $|G_{E}/G_{M}|$ determined precisely and lineshape of $|G_{E}|$ obtained for the first time. The recent results of neutron EMFFs at BESIII~\cite{BESIII:2021tbq} show great improvement comparing with
previous experiments, where the EMFFs are obtained at 18 energy points with the best precision achieving 8\%. A discrepancy is
observed in the cross section of $e^{+}e^{-}\to n\bar{n}$ to previous results and the long standing proton-neutron coupling puzzle is solved. Besides, an oscillation feature is observed in the residual
effective EMFF lineshape, with the same frequency but orthogonal phase to that of proton.
Cross section of various hyperon pair are studied from their thresholds. An anomalous enhancement behavior 
on the  $e^{+}e^{-}\to\Lambda\bar{\Lambda}$ cross section is observed at $\sqrt{s}=2.2324$~GeV~\cite{BESIII:2017hyw} where the non-vanish
cross section of neutral baryon pair near threshold is differing from pQCD prediction. Similar enhancement around $\sqrt{s}=2.2324$~GeV 
is observed in the cross section of  $e^{+}e^{-}\to KKKK$~\cite{BESIII:2019ebn}. 
The Born cross sections of $e^{+}e^{-}\to\Sigma\bar{\Sigma}$ are measured from their thresholds to 3.02~GeV~\cite{BESIII:2020uqk,BESIII:2021rkn}, where the lineshapes
can be well described by the pQCD-motivated function. However, a constant asymmetry in the cross sections is observed among the
$\Sigma$ isospin triplets. Similar study is performed on  $e^{+}e^{-}\to\Xi\bar{\Xi}$~\cite{BESIII:2021aer,BESIII:2020ktn} and the ratio of the Born cross sections
for the processes is found to be consistent with 1.

There are rich vector resonances around 2.0~GeV observed experimentally, interpreted as excited $\rho$, $\omega$ and $\phi$ resonances,
where the $\phi(2170)$ has drawn  a lot attention theoretically since it could be an exotic state candidate. 
Due to its controversial property, more experimental  researches are necessary.   
At BESIII, the lineshapes of $e^{+}e^{-}\to\phi\eta'$~\cite{BESIII:2020gnc} and $\phi\eta$~\cite{BESIII:2021bjn} are measured from 2.0 to 3.08 GeV.
A resonance structure are observed in both processes around 2.175~GeV, and it could be interpreted as a $\phi(2170)$   
due to rich $s\bar{s}$ component. 
Similar resonance structures are observed in $e^{+}e^{-}\to K^{+}K^{-}$~\cite{BESIII:2018ldc} and $K_{S}K_{L}$~\cite{BESIII:2021yam} lineshapes.
However, the mass of the resonance from the two processes are  different with average PDG value of $\phi(2170)$. 
Multiple lineshapes of intermediate state $e^{+}e^{-}\to K K_{1}(1400)$, $K K(1460)$, $K^{*}(892)K^{*}(892)$
and $K K_{1}(1270)$
are obtained by a partial wave analysis of $e^{+}e^{-}\to K^{+}K^{-}\pi^{0}\pi^{0}$~\cite{BESIII:2020vtu} at 10 energy points.
The structures observed in these intermediated processes are fitted  simultaneously and the partial width of each process is determined.
These results provide essential input to understand the nature of $\phi(2170)$.
The lineshapes of $e^{+}e^{-}\to\omega\pi^{0}$~\cite{BESIII:2020xmw} and $\eta'\pi^{+}\pi^{-}$~\cite{BESIII:2020kpr} are studied and the structures observed could be
excited $\rho$ resonances, such as $\rho(2000)$ or $\rho(2150)$.

\newpage

\subsection[Studies of LO and NLO FSR in the BABAR ISR $\mu^+\mu^-/\pi^+\pi^-/K^+K^-$ analyses]{Studies of LO and NLO FSR in the BABAR ISR\\ $\boldsymbol{\mu^+\mu^-/\pi^+\pi^-/K^+K^-}$ analyses}
\addtocontents{toc}{\hspace{2cm}{\sl M.~Davier}\par}

\vspace{5mm} 

Michel Davier

\vspace{5mm}

\noindent
IJCLab, Paris-Saclay University, Orsay, France
 \\
\vspace{5mm}


In this short abstract we recall the results of two original studies 
performed in the BABAR ISR experimental program, both dealing with FSR from 
muons, charged pions and kaons. Whereas FSR can be well predicted by QED for 
muons, the situation with hadrons is different, necessitating specific models. 
So far the simplest theoretical framework is scalar QED assuming a pointlike 
behaviour for the hadrons. However, for both LO and NLO, the observed FSR rate 
deviates from this simple model as shown by BABAR.
The neat idea of measuring the $e^+e^-\rightarrow {\rm hadrons}$ through the 
radiative process $e^+e^-\rightarrow \gamma ~{\rm  hadrons}$ assumes that the 
photon originates from ISR at LO. This is however not true for $\mu^+\mu^-$
production as the radiation could occur as well in the final state. In BABAR
experimental conditions at a CM energy of 10.58 GeV, the muon LO FSR amount 
remains at the 1\% level, at least for $\mu^+\mu^-$ invariant masses less 
than 1 GeV. It is expected to be much smaller for pions due to their composite 
nature.
 
A dedicated study~\cite{BaBar:2015onb} was aimed at
determining the LO FSR contributions for both $\mu^+\mu^-\gamma$ and 
$\pi^+\pi^-\gamma$ by measuring the ISR-FSR interference using a specific 
feature of the charge asymmetry in these processes. The FSR amplitude is 
found to be consistent with QED for muons. For pions the result disagrees with
the simple model, as used in Phokhara based on the pion form factor and FSR from 
point-like pions. However it agrees with a more realistic model where the photon 
is radiated from the produced quark pair at large $Q^2$ 
followed by a recombination of the quarks
into a pair of pions, dominated by $f_0(600)$ and $f_2(1270)$. Neglecting the 
$|\text{FSR}|^2$ contribution to the assumed-ISR cross section produces a bias for 
$a_\mu^\text{HVP, LO}$ of $(2.6\pm 1.2).10^{-11}$, which is negligible compared 
to the present uncertainty of $40.10^{-11}$ of the world-average HVP 
prediction~\cite{Aoyama:2020ynm}.

Taking correctly into account radiative corrections is an important aspect of 
$e^+e^-$ measurements. In particular FSR contributions have to be included when
evaluating the HVP dispersion relation. In most experiments this is done by 
adding the theoretical expectation to the LO measurement. The BABAR analyses 
on $\mu^+\mu^-\gamma(\gamma)$, 
$\pi^+\pi^-\gamma(\gamma)$~\cite{BaBar:2009wpw,BaBar:2012bdw}, 
and $K^+K^-\gamma(\gamma)$~\cite{BaBar:2013jqz} have proceeded differently by selecting 
data including the contributions of NLO photons through kinematical
fits allowing for one additional photon, either ISR or FSR. Furthermore the 
$\chi^2$ cuts are chosen to be extremely loose to keep a very high selection 
efficiency which is further measured on data. Measuring the FSR contribution
allows one to compare with the expectation using the simple scalar-QED model.
Among events with a detected and fitted additional photon, the BABAR analysis is 
able to separate the contributions from FSR and large-angle ISR photons. Results
are given for photon energies above 200 MeV and extending to a few GeV. In these 
conditions the ratio between the FSR fractions in data and in simulation is found 
to be 0.96$\pm$0.06 for $\mu^+\mu^-\gamma$, thus in agreement with NLO QED. The 
corresponding result for $\pi^+\pi^-\gamma$, 1.21$\pm$0.05, shows an excess over 
the scalar-QED prediction used in the Phokhara generator. The impact on the 
loose BABAR selection efficiency is small (0.6$\pm$0.2 per mil) and corrected for.
Measurements considering only LO photons without direct access to FSR would be 
affected more significantly.

\newpage

\subsection[Experimental asymmetry in CMD3 $2\pi$ data vs prediction]{Experimental asymmetry in CMD3 $\boldsymbol{2\pi}$ data vs prediction}
\addtocontents{toc}{\hspace{2cm}{\sl F.~Ignatov}\par}

\vspace{5mm} 

Fedor Ignatov

\vspace{5mm}

\noindent
Budker Institute of Nuclear Physics, SB RAS, Novosibirsk, 630090, Russia 
 \\

\vspace{5mm}

\begin{figure}[t]
\centering
\includegraphics[width=.57\linewidth]{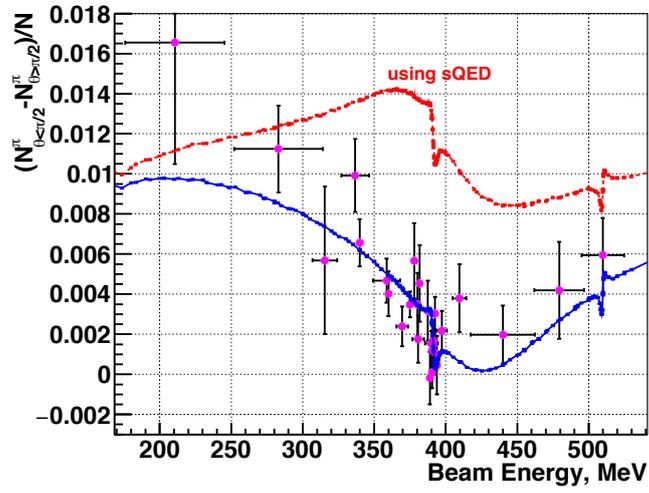}
\caption{The measured asymmetry in $2\pi$ at the CMD-3 in comparison with
  prediction based on commonly used sQED approach (red dotted line),
  and proposed above sQED calculation (blue line).}
\label{asym} 
\end{figure}

The dominant contribution to production of hadrons  in the energy range 
$\sqrt{s}<1$~GeV comes from the $e^+e^-\to\pi^+\pi^-$ mode. This channel gives the main contribution
to the hadronic term and overall theoretical precision of
the anomalous magnetic moment of the muon $g-2$ \cite{Aoyama:2020ynm}.
It is the most challenging channel because of a high-precision requirement,
 that must have a systematic precision of 0.2\% to be competitive with precision of new $g-2$ experiments and physics at future
electron--positron colliders.
The crucial pieces of analysis includes
stable particle separation, precise fiducial volume determination,
 theoretical precision of radiative corrections, etc. 
The calculated radiative corrections for the 2$\pi$ channel are based on different model
assumptions and the accuracy of this can affect different aspects of
channels analysis itself. The comparison of an experimental asymmetry
with predictions can be sensitive test for such models.
The charge asymmetry in the 2$\pi$ was studied with the
CMD-3 detector~\cite{Khazin:2008zz}. Understanding of the observed angular distributions is important for
investigation and controlling of systematic effects in the fiducial volume determination.
The measured asymmetry is defined as the difference of
detected number of events to the forward and backward region of the
detector: $A=(N(\theta<\pi/2)-N(\theta>\pi/2))/N$. This numbers was corrected
for all detectors effects like inefficiencies, resolution smearing,
etc, so the $A$ can be directly compared to predictions at the MC generator level.
The obtained result is shown in the Fig.~\ref{asym}. The points
correspond to the overall statistic collected at this moment with the
CMD3 detector at $\sqrt{s}<1$ GeV.

The dotted line correspond to the prediction based on the commonly used sQED
approach with the point-like pion assumption for radiative correction calculation~\cite{Arbuzov:2020foj}.
The observed discrepancy with experimental data has level about 1\% in the
$\rho$-peak region with $\pm$0.022\% statistical precision. Looking on
the dependency of asymmetry  
versus the invariant mass of pions gives hints that such
discrepancy comes from the soft and virtual parts of the radiative
corrections. One of the critical points for this parts comes from
applicability of the sQED assumption
for the box diagrams. More proper way will be to include the pion formfactor to each
vertex. Such calculation~\cite{romanlee} was performed and it gives
significant enhancement of the virtual corrections. The proposed
improvement gives good description of the experimental data as shown by the
blue line at the Fig.~\ref{asym}, and they are consistent at statistical precision.
This result show importance of the better models for some part of
radiative corrections for the 2$\pi$ channel, as they can gives sizable
effect both in the charge-odd and in some cases in the charge-even parts. It should be cross-checked with
re-calculation for the ISR measurement, where already unaccounted effects
at the NNLO order, even with sQED assumption,
was shown in the paper for the Phokhara generator~\cite{Campanario:2019mjh}.

\clearpage

\section{Report on the Precision Standard Model Database for the STRONG2020 Project}
\addtocontents{toc}{\hspace{2cm}{\sl A. Driutti}\par}

\vspace{5mm} A. Driutti, on behalf of the PrecisionSM DB group

\vspace{5mm}

\noindent
University and INFN Pisa
 \\
\vspace{5mm}

The PrecisionSM database is one of the specific objectives of Strong2020, a European joint research initiative that groups together researchers from different scientific frontiers (low energy, high energy, instrumentation and infrastructures) with the broad goal of study strong interactions and  develop applications for beyond fundamental physics~\cite{Refstrong2020}. 

The plan for implementing the precisionSM database is summarized in the following steps. The first step consists in collecting, with the help of experts, the list of published precision low energy data from various experiments. Each of these measurements is then catalogued and made available in a public repository called HEPData~\cite{RefHEPData}. A coordinator, usually a point-of-contact person designated by each experiment or project, uploads the data in the repository and appoints a reviewer to perform a cross-reference validation. Validated data are then made public and indexed in an easy, transparent and accessible way through the PrecisionSM database webpage~\cite{RefPrecisionSM}.

At present, we are collecting measurements from the $e^+e^- \rightarrow \pi^+\pi^-$ channel, which is an important channel for computing the muon $g-2$ hadronic vacuum polarization term, from the following experiments~\cite{RefBaBar,RefBESIII,RefKLOE,RefNovosibirsk,Refold}: BaBar (SLAC, USA); Beijing Spectrometer (BESIII, China); KLOE (LNF, Italy); SND, TOF, OLYA, VEPP, CDM, CMD-2 and CMD-3 (Novosibirsk, Russia); MEA and BCF (Adone LNF, Italy), CLEO (Cornell Electron Storage Ring, USA). These datasets are in the process to be uploaded and validated in HEPData. Measurements details are also being collected in the PrecisionSM website. The plan for the future is to catalogue measurements for other $e^+e^- \rightarrow \text{hadrons}$ channels as well as develop tools to perform data downloading and simple elaboration, e.g., comparative plots.

\newpage


\section{Radiative Corrections and MC generators for time-like processes}

\subsection{MCGPJ and ReneSANCe Monte Carlo event generators: status and perspectives}
\addtocontents{toc}{\hspace{2cm}{\sl A.~Arbuzov}\par}

\vspace{5mm} Andrej Arbuzov

\vspace{5mm}

\noindent
Joint Institute for Nuclear Research
\\
Joliot-Curie str. 6, Dubna, 141980, Russia
\vspace{5mm}

The substantial progress both in theoretical and experimental
studies of low-energy lepton interactions motivates us to
reconsider the status of existing Monte Carlo event generators
which are used in these studies and to think about creation
of new advanced codes. In particular, that is motivated by
requests of forthcoming new experiments like MUonE~\cite{Abbiendi:2016xup} 
at CERN and Super Charm-Tau Factory 
in Novosibirsk~\cite{Charm-TauFactory:2013cnj}.\footnote{The Novosibirsk project can be realized in Sarov 
and there is a similar project in China.}

{\tt MCGPJ}~\cite{Arbuzov:2005pt} is a Monte Carlo event generator wchich allows 
to simulate a number of electron--positron annihilation processes at
energies up to a few GeV. This code has a universal treatment of higher-order 
collinear initial state radiation (ISR) and complete one-loop QED corrections.
That allows to reach about 0.2\% of even better for sufficiently inclusive
(IR-safe) observables. But the coode looses precision in description of differential 
distributions especially at extreme kinematics. That is because of approximate 
treatment of kinematics of events with multi-photon emission.
There are two general way to improve the code precision: 
1) implement an advance matching of NLO QED corrections; 
2) realize complete ${\mathcal O}(\alpha^2)$ corrections with some 
(approximate) treatment of higher-order leading contributions. 
Implementation of either option might require creation of a completely 
new code. Note that two-loop QED radiative corrections nowadays become
available for more and more processes.

In any case, an accurate implementation of one-loop corrections is 
mandatory. Recently a new Monte Carlo event generator 
{\tt ReneSANCe}~\cite{Sadykov:2020any} was created. Now it can simulate
several processes of electron--positron annihilation, Bhabha scattering,
and muon--electron scattering at a wide range of energies with taking into 
account the complete ${\mathcal O}(\alpha)$ electroweak radiative 
corrections. Moreover, this code allows particle polarization which
is relevant for future experiments. In particular, the updated
version of {\tt ReneSANCe}~\cite{Arbuzov:2021oxs}
can describe scattering of polarized muons off electrons for the conditions
of the MUonE experiment while other existing codes typically work only
for unpolarized case of this process.

Future plans in development of the {\tt ReneSANCe} code are: 
i) tuned comparisons with other codes for processes relevant for
experiments, e.g., Bhabha and $\mu e$ scattering in particular; 
ii) implementation of other processes ($e^+e^-\to\gamma\gamma$, 
$e^+e^-\to \gamma Z$, $e^-e^-\to e^-e^-$ etc.); 
iii) implementation of known two-loop results; 
iv) matching with QED parton showers.

\newpage

\subsection{The Monte Carlo event generator Phokhara}
\addtocontents{toc}{\hspace{2cm}{\sl H.~Czy\.z}\par}
\vspace{5mm} Henryk Czy\.z

\vspace{5mm}

\noindent
Institute of Physics, University of Silesia, Katowice, Poland
 \\
\vspace{5mm}

The buildout of the Monte Carlo event generator Phokhara started with the paper
\cite{Rodrigo:2001kf} and continued up to Ref.~\cite{Campanario:2019mjh}. During
about twenty years of the code development the group concentrated
 on two aspects vital
for a good description of the hadron production in $e^+e^-$ annihilation: the proper
modeling of interactions between photons and hadrons and calculations of
radiative corrections. For each hadronic final state a dedicated sampling method
was constructed allowing for a fast Monte Carlo generation. The three most important
papers on radiative corrections, which are implemented in the Phokhara generator
are Refs.~\cite{Rodrigo:2001jr,Campanario:2013uea,Campanario:2019mjh}. In the first one
the universal, for all hadronic states, initial state radiative corrections were calculated at
Next to Leading Order (NLO). In the second one complete radiative corrections at NLO
 for
 the reaction $e^+e^-\to \mu^+\mu^-\gamma$ were determined. While in the third one
  complete radiative corrections at NLO
 for
 the reaction $e^+e^-\to \pi^+\pi^-\gamma$ were calculated.
 The code simulates reactions  $e^+e^-\to R (\gamma)$. Where
 R is one of the following states: $\mu^+\mu^-$, $\pi^+\pi^-$, $2\pi^0\pi^+\pi^-$,
 $2\pi^+2\pi^-$, $p\bar{p}$, $n\bar{n}$, $K^+K^-$, $K^0\bar K^0$,
 $\Lambda (\to \pi^- p) \bar \Lambda (\to \pi^+ \bar p)$, $\eta\pi^+\pi^-$,
 $\pi^0 \gamma$, $\eta \gamma$, $\eta^{\prime} \gamma$,
 $\chi_{c_1}\to J/\psi(\to\mu^+\mu^-)\gamma$,
 $\chi_{c_2}\to J/\psi(\to\mu^+\mu^-)\gamma$. In the last two the QED background is also included \cite{Czyz:2016xvc}. Models of various hadronic currents used
 in simulations where developed in Refs.~\cite{Czyz:2010hj}
 ($\pi^+\pi^-$,$K^+K^-$, $K^0\bar K^0$), \cite{Czyz:2000wh,Czyz:2008kw}
 ($2\pi^0\pi^+\pi^-$,
 $2\pi^+2\pi^-$), \cite{Czyz:2014sha}($p\bar{p}$, $n\bar{n}$),
 \cite{Czyz:2007wi} ($\Lambda (\to \pi^- p) \bar \Lambda (\to \pi^+ \bar p)$, $\eta\pi^+\pi^-$), \cite{Czyz:2012nq,Czyz:2017veo}
 ($\pi^0 \gamma$, $\eta \gamma$, $\eta^{\prime} \gamma$) and
  \cite{Czyz:2016xvc} ($\chi_{c_1}\to J/\psi(\to\mu^+\mu^-)\gamma$,
 $\chi_{c_2}\to J/\psi(\to\mu^+\mu^-)\gamma$).
 
 \newpage
 
\subsection{KKMCee/BHLUMI/BHWIDE MC generators: status and prospects}
\addtocontents{toc}{\hspace{2cm}{\sl S.~Jadach}\par}
\vspace{5mm} 
Stanis\l{}aw Jadach

\vspace{5mm}

\noindent
Institute of Nuclear Physics PAN, Krak\'ow, Poland \\
\vspace{5mm}

KKMC is the MC event generator for the process
$ e^+e^- \to f\bar{f}+n\gamma$,
$f = \mu,\tau,\nu,u,d,s,c,b$, $n = 0,1,2...\infty$.
Published version 4.13 
with F77 code description and user guide (manual)
of Ref.~\cite{Jadach:1999vf} is to be cited.
Ref.~\cite{Jadach:2000ir} describes
physics content, CEEX exponentiation of QED matrix element.
KKMC is interfaced with TAUOLA \cite{Jadach:1993hs}, 
PHOTOS~\cite{Barberio:1993qi} and with electroweak library DIZET~\cite{Bardin:1989tq}.
It was used in data analysis of all four LEP collaborations.
KKMC is unique because its implements several features not seen in any other MC:
(a) soft photon resummation of real and virtual corrections to infinite order,
CEEX type matrix element~\cite{Jadach:2000ir}
(b) exact multiphoton phase space
(c) full control of spin polarization (longitudinal and transverse) of all particles
(d) narrow resonance effects
(e) initial-final state and all other interferences
(f) complete photonic 2-nd order QED corrections
(g) complete 1-st order EW corrections.
(h) parton shower hadronization of $q\bar{q}$ pairs.
KKMC is capable to calculate QED effects with $0.01\%$ precision,
as shown in Ref.~\cite{Jadach:2018lwm}.
The 10ppm precision needed for FCCee will require improved CEEX matrix element.
Public f77 version interfaced to FCCee software is available on GITHUB server
\url{https://github.com/KrakowHEPSoft/KKMCee/releases/tag/v4.32.01}.%

C++ version of KKMC is already there, to be published soon.
Complete documentation of KKMC and older source code versions are available on
\url{http://192.245.169.66:8000/FCCeeMC/wiki/kkmc}.%

The classic version of BHLUMI Monte Carlo for low angle Bhabha process
of Ref.~\cite{Jadach:1996is} is maintained in a minimal way.
Version which compiles under modern f77 linux compilers 
can be found on
\url{http://192.245.169.66:8000/FCCeeMC/wiki/bhlumi},%
together with its documentation.
Further planned development of BHLUMI for FCCee is outlined
in Ref.~\cite{Jadach:2018jjo}.

BHWIDE Monte Carlo for wide angle Bhabha 
of Ref.~\cite{Jadach:1995nk}, 
compiling under linux f77 compilers is available at
\url{http://placzek.web.cern.ch/placzek/}.
Both BHWIDE and BHLUMI should be incorporated into KKMC.

\newpage

\subsection[The $\tau$ lepton Monte Carlo Event Generation --
imprinting  New Physics models with exotic scalar or vector states into simulation samples]{The $\boldsymbol{\tau}$ lepton Monte Carlo Event Generation -- \\
imprinting  New Physics models with exotic scalar or vector states into simulation samples}
\addtocontents{toc}{\hspace{2cm}{\sl Z.~Was\par}}

\vspace{5mm}

{Sw. Banerjee$^a$, D. Biswas$^a$, T. Przedzinski$^b$, Z. Was$^{\dagger,c}$}

\vspace{5mm}

\noindent
{\em $^a$ University of Louisville, Louisville, Kentucky, 40292, USA}\\
{\em $^b$ Institute of Physics, Jagellonian University, 30-348 Krakow, Lojasiewicza 11, Poland }\\
{\em $^c$ Institute of Nuclear Physics, Polish Academy of Sciences, PL-31342 Krakow}

\vspace{2mm}

\noindent$^{\dagger}$ Speaker

\vspace{5mm}

The Monte Carlo for lepton pair production and $\tau$ decays consist
of {\tt KKMC} for lepton pair production, {\tt tauola} for $\tau$ lepton decays and
{\tt photos} for radiative corrections in decays.

An effort for adaptation of the system for precision data being collected
at the {\tt Belle II} experiment included simulation of  additional light lepton
pairs. Extension to processes where lepton pair is produced through narrow resonances,
like dark photon or dark scalar ($\phi$) resonances, was straight forward.

Modified  programs versions are available in stand-alone format from gitlab repository or through the {\tt basf2} system of
Belle II software. It was explained recently  during the International Workshop on Tau Lepton Physics September, 2021, Bloomington IN. Now we concentrate on
simulations for $\phi$ resonance, a hypothetical object which could be responsible for anomalous moment $g-2$ in $Z-\tau-\tau$ interactions through virtual contributions.

For details see Refs.~\cite{Banerjee:2021hke,Banerjee:2021rtn}

\newpage

\subsection{Discrepancies between current MC generators}
\addtocontents{toc}{\hspace{2cm}{\sl F.~Ignatov}\par}

\vspace{5mm} Fedor Ignatov

\vspace{5mm}

\noindent
Budker Institute of Nuclear Physics, SB RAS, Novosibirsk, 630090, Russia 
 \\
\vspace{5mm}

MC generators is one of the main important tool to perform physics study.
Current level of analysis require knowledge of
integrated cross-section of main channels $e^+e^- \rightarrow e^+e^-,
\pi^+\pi^-, \mu^+\mu^-$ with precision $\leq0.1\%$.
Current and future experiment will have increased collected luminosity,
which allow to do precise analysis based on some differential
quantities. This require a precise prediction not only for the total
cross-section in an acceptance cuts, but also it is require a good knowledge on shapes of distributions.

In case of the bhabha scattering process, several MC generators is
available with 0.1-0.5\% precision:
BaBaYaga@NLO~\cite{Balossini:2006wc} - was developed for the KLOE
experiment and now widely used in many others,
MCGPJ~\cite{Arbuzov:2005pt} - mainly used in the Novosibirsk $e^+e^-$
experiments, BHWIDE~\cite{Jadach:1995nk} - was born in the LEP era.
All this generators include exact NLO + higher order terms in some
approximation (the parton shower approach, the YFS exponentiation
method, collinear structure functions, etc). They all are
well consistent in the integrated cross-section at level better than
0.1\% (having mostly same $\alpha$ order inclusion). Also they can be
consistent on some of final particles kinematic distributions, but
they still have discrepancies in some specific selections on 2D
distributions. For example if one lepton is selected as
$0.3<P^{+}/E_\text{beam}< 0.45$ (with additional detector specific
colliniarity cuts), then the other charge momentum distribution will
have inconsistency at $\sim10\%$ level between MCGPJ vs
BabaYaga@NLO vs Bhwide.
The differential cross section over polar angle
are inconsistent at $0.1-0.2\%$ level between them, and
this is already the sensitive number for the asymmetry study in $2\pi$ with
the CMD3 detector. Other new generators appeared recently:
the ReneSANCe generator~\cite{Sadykov:2020any}
are under development now by Dubna group. While it was still
missed some parts with leading log corrections at the moment of the
workshop, but this group have plans to include NNLO in future and they keep possibility
to have polarization of colliding beams from initial stages (which was out of
consideration for mostly all generators). The formulas for the NNLO of
the bhabha process was known for a long time, but it is just recently appeared first tool
with inclusion of them. This was implemented at fixed order in
the McMule framework~\cite{Banerjee:2021mty}, which do the matrix element
integration in desired acceptance cuts.

For the $e^+e^- \rightarrow \pi^+\pi^-$ channel, it is seen lack of
available precise generators for comparison. We have the PHOKHARA
generator with 0.5\% claimed systematic accuracy.
It was developed for the ISR process study,
where complete set of NLO to $e^+e^- \rightarrow \pi^+\pi^-\gamma$ was
implemented in the latest version~\cite{Campanario:2019mjh}.
But this generator has limited precision
for scanned mode (without requirement on additional photon).
And we have the MCGPJ generator with 0.2\% claimed accuracy, which includes exact
NLO to $e^+e^- \rightarrow \pi^+\pi^-$ with point-like pion and
additional ISR jets along beam with help of the structure functions.
But this generator is not supposed to
be used for ISR studies. Both generators have different region of applicability.

For the $e^+e^- \rightarrow \mu^+\mu^-$ channel, many
generators are available at same NLO order or some parts above: BaBaYaga@NLO, MCGPJ,
Phokhara, KKMC~\cite{Jadach:1999vf}, and other more precise will appeared soon in
light of new outstanding the MuOnE experiment.
If to do comparison of current generators, it is seen that 
an exact muon mass term in the FSR virtual
correction is commonly missed in most of them, which should be
present according to the analytical total cross section
calculation~\cite{Bystritskiy:2005ib}. 
It gives an additional uncertainty which behaves as $1/\beta_l$ with $>0.2\%$ at $\beta_l<0.88$.
This energy range is in use with VEPP-2000 experiments,
and it should be important also for the $\tau\tau$ production at threshold.

Another important ingredient of all generators is the vacuum
polarization parameterization. Most up to day compilations come from
the KNT group~\cite{Keshavarzi:2018mgv}, Fred
Jegerlehner~\cite{Jegerlehner:2019lxt}, Novosibirsk~\cite{Ignatov:2008bfz}.
All of them well consistent at 0.05-0.1\% level outside of narrow
resonances. But it is the different situation at narrow resonances, for example for
near $\phi$ region they have statistical inconsistency at $\sim0.5\%$ with
up to 1.5-2.\% when the resonances was treated as ``dressed'' one in input
(while should be ``bare'').
It will be worthwhile to look on comparison between different
compilations if someone are doing analysis on narrow resonances
with below this mentioned precision.

\newpage

\subsection[Dispersive approach to isospin-breaking corrections to $e^+e^- \to \pi^+ \pi^-$ and $\pi^+\pi^- \to \pi^+ \pi^-$]{Dispersive approach to isospin-breaking corrections to\\ $\boldsymbol{e^+e^- \to \pi^+ \pi^-}$ and $\boldsymbol{\pi^+\pi^- \to \pi^+ \pi^-}$}
\addtocontents{toc}{\hspace{2cm}{\sl G.~Colangelo}\par}

\vspace{5mm} Gilberto Colangelo

\vspace{5mm}

\noindent
Albert Einstein Center for Fundamental Physics,\\ Institute for Theoretical Physics, University of Bern, \\Sidlerstrasse 5, 3012 Bern, Switzerland
 \\

\vspace{5mm}

In this talk I have reported about work in progress done in collaboration
with Joachim Monnard and Jacobo Ruiz de Elvira which aims to
calculate the isospin breaking corrections to $e^+e^- \to
  \pi^+ \pi^-$ and $\pi^+\pi^- \to \pi^+ \pi^-$ in a dispersive
  way~\cite{RefColangelo}. Some preliminary results have already been published in
  the PhD thesis of Joachim Monnard~\cite{JMPhDThesis}.

We take into account all three sources of isospin breaking, namely :
1. $m_u-m_d$, 2. $M_{\pi^+}-M_{\pi^0}$ and 3. photonic radiative
corrections. Note that although the pion mass difference is generated
mainly by photonic effects it is possible to cleanly separate 2. from 3.,
see discussion in Ref.~\cite{Colangelo:2008sm}. Effects generated by
$m_u-m_d$ are essentially related to $\rho$-$\omega$ mixing, which is
already included in dispersive analyses of the pion vector form
factor $F_V^\pi$~\cite{Colangelo:2018mtw}. The quark mass difference also
generates a $\pi^0$-$\eta$ mixing, which however only indirectly affects
$F^V_\pi$.  

Also effects related to the pion mass difference indirectly influence the
pion vector form factor, since in the main channel only charged pions are
exchanged. In the $\pi \pi$ scattering amplitude, however, they are
important close to threshold. They can be taken into account by rewriting
the Roy equations in the charge basis and using physical masses, in
particular in the lower boundaries of the dispersive integrals. The effect
is small and can be taken into account by iterating a few times the Roy
equations in the charge basis and with the shifted integration boundaries.

The evaluation of photonic radiative corrections is more complicated. A
full-fledged dispersive approach is probably out of reach, but as long as
one is interested in the region below 1 GeV, it makes sense to adopt the
approximation which was the key to the dispersive treatment of hadronic
light-by-light~\cite{Colangelo:2017fiz}, namely to only consider up to
two-pion (plus possibly a photon) intermediate states. This approximation
leads to a limited number of dispersive diagrams one has to consider
and correspondingly to the number of subamplitudes. Solving the dispersion
relations that one obtains in this way is nontrivial and requires to
numerically solve a new kind of integral equations. Preliminary results in
this direction have been presented in the PhD Thesis already
mentioned~\cite{JMPhDThesis}. These indicate that although the energy
dependence of the radiatively corrected cross section is more complicated
than so far evaluated (see for example Ref.~\cite{Hoefer:2001mx}), the size of
the corrections has been estimated in the right ballpark.

With the increasing precision requirements for the evaluation of the HVP
contribution to the $(g-2)_\mu$ it is important to correctly assess the
size as well as the energy dependence of these radiative correction to the
$e^+e^- \to \pi^+ \pi^-$ cross section: this is what our work in
progress~\cite{RefColangelo} is aiming for.

\newpage 

\subsection[Perspectives from theory on $e^+e^-\to 2\pi$ and $e^+e^-\to 3\pi$]{Perspectives from theory on $\boldsymbol{e^+e^-\to 2\pi}$ and $\boldsymbol{e^+e^-\to 3\pi}$ }
\addtocontents{toc}{\hspace{2cm}{\sl M.~Hoferichter}\par}

\vspace{5mm} Martin Hoferichter

\vspace{5mm}
\noindent
Albert Einstein Center for Fundamental Physics,\\ Institute for Theoretical Physics, University of Bern, \\Sidlerstrasse 5, 3012 Bern, Switzerland
 \\

\vspace{5mm}

In the talk we brought up the question what can actually be said in a rigorous way about the cross sections for $e^+e^-\to 2\pi$ and $e^+e^-\to 3\pi$, given that vector meson dominance cannot serve as a quantitative guide at the subpercent level of precision required for $(g-2)_\mu$. Two classes of model-independent constraints exist~\cite{Colangelo:2018mtw,Hoferichter:2019mqg,Hoid:2020xjs,Colangelo:2020lcg}: (i) mass, width, and mixing parameters of $\omega$ and $\phi$ on the real axis can be interpreted as pole parameters and residues, since corrections from the analytic continuation are sufficiently suppressed by the small widths, and thus need to be reaction independent; (ii) the line shape of the resonances is determined by their main decay channels, via amplitudes that in turn are subject to constraints from analyticity, unitarity, and crossing symmetry. 

We then discussed the status of the $\omega$ and $\phi$ parameters, pointing out issues with the average for the $\omega$ mass in Ref.~\cite{ParticleDataGroup:2020ssz} (regarding the treatment of hadronic $\bar p p$ reactions and an unphysical phase in $e^+e^-\to\pi^0\gamma$, both of which drive a large scale factor). Once these are resolved, no scale factor is required and good consistency emerges between $e^+e^-\to 3\pi,\pi^0\gamma, \bar K K$~\cite{Hoid:2020xjs}. This picture changes when the new data sets for $e^+e^-\to 3\pi$ are considered, whose resonance masses differ slightly~\cite{BABAR:2021cde} or even substantially~\cite{BESIII:2019gjz} from previous measurements, potentially affecting the combination of data sets and vacuum-polarization corrections. In addition, the $\omega$ parameters need to be consistent with $e^+e^-\to2\pi$, where they enter as a resonance-enhanced isospin-breaking effect, but it has been known for a while that the $\omega$ mass tends to come out substantially smaller than in $e^+e^-\to 3\pi$ unless a large phase $\delta_\epsilon\simeq 10^{\circ}$ in the mixing parameter is permitted~\cite{BaBar:2012bdw}. However, such a large phase cannot be justified from radiative intermediate states $\pi^0\gamma, \pi\pi\gamma, \eta\gamma,\ldots$, which do produce imaginary parts, but only of the size $\delta_\epsilon \lesssim 4^\circ$. Moreover, preferences among the different experiments vary widely, with the data from Refs.~\cite{BaBar:2012bdw,KLOE-2:2017fda,SND:2020nwa} suggesting $\delta_\epsilon\simeq 1^\circ, 6^\circ, 10^\circ$, respectively, and the fit quality to Ref.~\cite{SND:2020nwa} depending crucially on allowing such a large phase, while other fits only improve moderately when introducing a non-vanishing $\delta_\epsilon$.   

All of this suggests that we do not yet have a consistent picture even for the simplest hadronic reactions, but given the model-independent constraints available in these cases, we believe that the corresponding consistency checks are necessary to be able to claim an HVP precision at few-permil level. In particular, insisting on reaction-independent $\omega$ and $\phi$ parameters is of paramount importance already for the direct integration of data sets, since the consistent removal of vacuum polarization depends critically on the resonance positions~\cite{Davier:2019can,Keshavarzi:2019abf}. We suspect that more information is required on energy resolution and/or calibration uncertainties to move forward. 

\newpage 

\subsection{Mixed leptonic and hadronic corrections to the anomalous magnetic moment of the muon}
\addtocontents{toc}{\hspace{2cm}{\sl T.~Teubner}\par}

\vspace{5mm} Thomas Teubner

\vspace{5mm}

\noindent
Department of Mathematical Sciences, University of Liverpool,
Liverpool\ \ L69 3BX, U.K.
 \\
\vspace{5mm}

Estimates of mixed leptonic and hadronic contributions to $g-2$ of
the muon, which have so far not been included in the Standard Model
prediction, were presented. Following the discussion at the Strong 2020
meeting, these estimates were quantified in detail, see
Ref.~\cite{Hoferichter:2021wyj}, where we conclude that these
`double-bubble' NNLO corrections are negligible at the level of the
expected final precision of the Fermilab $g-2$ experiment.

\newpage


\section{Radiative Corrections and MC generators for space-like processes}

\subsection[Space-like method for hadronic vacuum polarization contributions to muon $g-2$]{Space-like method for hadronic vacuum polarization contributions to muon $\boldsymbol{g-2}$}
\addtocontents{toc}{\hspace{2cm}{\sl S.~Laporta}\par}

\vspace{5mm} Stefano Laporta

\vspace{5mm}

\noindent
Istituto Nazionale di Fisica Nucleare, Sezione di Padova, Padova,
Italy\\
Dipartimento di Fisica, Universit\`a di Bologna e 
Istituto Nazionale di Fisica Nucleare, Sezione di Bologna, Bologna
 \\

\vspace{5mm}

The hadronic vacuum polarization (HVP) contribution to the muon $g-2$
is traditionally computed via dispersive time-like integral
using hadronic production cross section in $e^{-}e^{+}$ annihilation.
In this talk we describe the results of 
a joint work with Elisa Balzani and Massimo
Passera~\cite{Balzani:2021del}.
First, we review the well-known space-like integral formula
for the LO contribution to HVP~\cite{Lautrup:1971jf}.
Then, we consider the NLO contributions to HVP (see figure):
\\
\centerline{
  \includegraphics[scale=0.4]{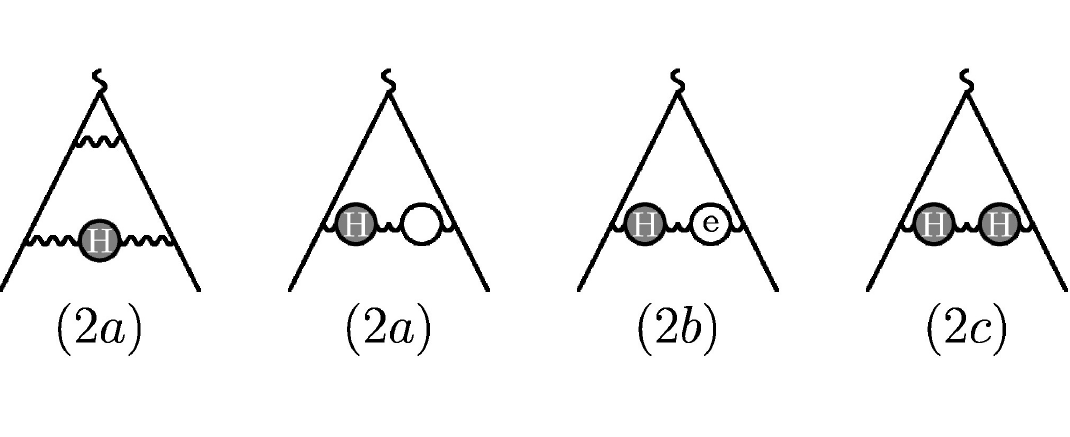}
}
Using the results of Ref.~\cite{Barbieri:1974nc},
we obtain the following time-like integral formulas for the NLO HVP
contributions of the classes of diagrams (a), (b)
and (c), respectively:
\begin{align*}
a_{\mu}^{\textrm{HVP}}(\textrm{NLO;a})&=\left(\frac{\alpha}{\pi}\right)^2
\int_{0}^{1}{dx}\;
 {\kappa}^{(4)}(x) \Delta\alpha_{\textrm{h}}(t(x))  \ ,
 \cr
a_{\mu}^{\textrm{HVP}}(\textrm{NLO;b})&=\frac{\alpha}{\pi}
\int_{0}^{1}{dx\;}
 {\kappa}^{(2)}(x) \Delta\alpha_{\textrm{h}}(t(x))
 \
  2\left(
  \Delta \alpha^{(2)}_e (t(x))
 +\Delta \alpha^{(2)}_{\tau} (t(x)) \right)
 \ ,
 \cr
a_{\mu}^{\textrm{HVP}}(\textrm{NLO;c})&=\frac{\alpha}{\pi}
\int_{0}^{1}{dx}\;
 {\kappa}^{(2)}(x) \left(\Delta\alpha_{\textrm{h}}(t(x))\right)^2
 \ ,
\end{align*}
where ${\kappa}^{(2)}$ and ${\kappa}^{(4)}$ are the exact space-like kernels 
${\kappa}^{(2)}(x)=1-x$,
${\kappa}^{(4)}(x)=
\frac{2(2-x)}{x(x-1)} F^{(4)}(x-1)  \ ,
$
\begin{align*}
F^{(4)}(y)=&
\tfrac{\left(-3 y^4-5 y^3-7 y^2-5 y-3\right) \left(2 \text{Li}_2(-y)+4
\text{Li}_2(y)+\ln (-y) \ln \left((1-y)^2 (y+1)\right)\right)}{6
   y^2}
\cr 
&
   +\tfrac{(y+1) \left(-y^3+7 y^2+8 y+6\right) \ln (y+1)}{12
   y^2}+\tfrac{\left(-7 y^4-8 y^3+8 y+7\right) \ln (1-y)}{12
   y^2}
\cr
&
   +\tfrac{23
      y^6-37 y^5+124 y^4-86 y^3-57 y^2+99 y+78}{72 (y-1)^2 y
      (y+1)}
      +\tfrac{\left(12 y^8-11 y^7-78 y^6+21 y^5+4 y^4-15 y^3+13
      y+6\right) \ln(-y)}{12 (y-1)^3 y (y+1)^2}\ ;
\end{align*}	 
$\Delta\alpha_h(t)=-\Pi_h(t)$ is (five-flavor) hadronic
contribution to the running of the effective fine structure constant in
the space-like region, and 
$\Pi_l^{(2)}(t)=-\Delta\alpha_{\textrm{l}}(t)$ is the renormalized QED
vacuum polarization function for the lepton $l$.

At last, we analyze the approximation to ${\kappa}^{(4)}(x)$
considered in Ref.~\cite{Chakraborty:2018iyb}. 
We show that the approximation gives raise to an error of $\sim 6 \% $  of the total NLO
contribution, and we describe some ways to improve it. 

\newpage 

\subsection{Status of MUonE}
\addtocontents{toc}{\hspace{2cm}{\sl U.~Marconi}\par}

\vspace{5mm} Umberto Marconi

\vspace{5mm}

\noindent
{INFN - Sezione di Bologna, Via Irnerio, 46, 40126 Bologna  -   Italy}
 \\
\vspace{5mm}

%
The MUonE experiment has been presented at CERN with the Letter-of-Intent~\cite{MUonE:LoI} in June 2019. 
MUonE aims at a new determination of the hadronic vacuum polarization (HVP) contribution to the muon $(g-2)_{\mu}$.
The HVP can be obtained through the effective electromagnetic coupling in the space-like region for low momentum transfer,
by a very precise determination of the shape of the differential muon--electron elastic cross section,
exploiting the scattering of 160 GeV muons on atomic electrons of a low-Z target.
The measurement can be performed at CERN's M2 muon beam. 
MUonE plans to determine the HVP with a goal precision of $\mathcal{O}$(10$^{-3}$).
In this purpose the shape of the differential cross section as a function of the scattering angles has to be measured to the remarkable precision of $\mathcal{O}$(10$^{-5}$).
The detector consists of 40 identical tracking stations, each equipped with a thin target (1.5 cm of Beryllium or an equivalent Carbon thickness) and three planes per orthogonal view of silicon strip detectors. 
A station has a length of 1 m and transverse dimensions of about 10 cm, resulting in a well suited  angular acceptance of 50 mrad.
An electromagnetic calorimeter and a muon detector located at the end of the detector are used for the PID, and to control the level of the background contamination surviving to the elastic selection.
It will be crucial to select elastic events with uniform efficiency, rejecting inelastic (e.g. nuclear interactions, pair production) or radiative ones. The expected precise correlation between the scattering angles will help in the selection.
The experiment has been designed to reach the required statistical precision in three years of data taking, with an integrated luminosity of $1.5 \times 10^{7}$ nb$^{-1}$.

The challenge of the experiment is to keep the systematic error at the level of the statistical precision. The main source of systematic uncertainty is related to the precision achievable in determining the mean beam energy. It translates into a very demanding precision in the determination of the longitudinal position of the sensors to 10 micron. 

Test beams have been successfully carried out in order to study the effects of multiple scattering induced by the target material the angular distributions, and to select a clean sample of elastic collisions of muons on atomic electrons.

For the final tracking detector MUonE plans to rely on the 2S modules developed by CMS for its upgrade. 
The 2S modules provide stubs at 40 MHz, i.e. hits detected in coincidence by paired sensors assembled in a module. The high counting rate capability of the modules guarantees a small pileup despite the expected high beam intensity of about 50 MHz. 
The DAQ system will rely on the Serenity readout board, which is capable of managing the expected high data throughput. A trigger system to reduce the data flux to the DAQ servers can be implemented exploiting on board powerful FPGAs.

First tests of the 2S modules, in collaboration with the CMS-TK tracking group are presently ongoing at the M2 muon beam line to evaluate hits efficiency and the spatial resolution.

\newpage 

\subsection{MUonE analysis status}
\addtocontents{toc}{\hspace{2cm}{\sl G.~Abbiendi}\par}

\vspace{5mm} Giovanni Abbiendi

\vspace{5mm}

\noindent
{INFN - Sezione di Bologna, Viale C. Berti Pichat, 6/2, 40127 Bologna  -   Italy}
 \\
\vspace{5mm}

%
The proposal of the MUonE experiment has been presented at CERN with the Letter-of-Intent~\cite{MUonE:LoI} in June 2019. The updated status of the project has been presented at this workshop, see previous contribution.
In this contribution the analysis strategy is reviewed.

The running of $\alpha_\text{QED}$ will be measured in MUonE, with the 160
GeV M2 muon beam at CERN SPS, from the shape of the differential cross section for elastic scattering of muons
on atomic electrons of fixed targets.
This precise measurement can be achieved by a template fit of the angular
distributions, by reweighting the effective QED coupling with a
convenient parameterisation (one-loop QED, or {\em Lepton-Like}, with only
2 parameters).
The expected distributions have been calculated by the MESMER Monte
Carlo at NLO \cite{Alacevich:2018vez}, which is an exact
$\mathcal{O}(\alpha)$ calculation including masses ($m_e$, $m_\mu$) and electroweak
contributions. Detector effects including multiple Coulomb
scattering and intrinsic resolution are parameterised and treated by a
fast simulation.
The fit can be done on the 1D distribution of either the electron or the muon scattering angle, or on the 2D
($\theta_e$, $\theta_\mu$) distribution, which is the most robust option. Actually, particle
identification is not strictly necessary.
By inserting the fitted parameterisation of $\Delta\alpha_\text{had}$ in
the master integral, the $a_\mu^\text{HVP, LO}$ can be obtained with
expected statistical accuracy of 0.35\%
for the MUonE nominal integrated luminosity, with negligible error from the fit method.

Prospects for a first physics run foreseen in 2022 with three tracking
stations have been assessed.
An integrated luminosity of about 5~pb$^{-1}$ is estimated with one week
of data taking under nominal beam conditions.
Such a data sample would yield $\mathcal{O}$(10$^{9}$)
events with scattered electron with energy greater than 1~GeV.
This should allow a measurement of the leptonic running with good precision, and could
even provide initial sensitivity to the hadronic running.

Preliminary studies of the main systematic effects showed that they
can be controlled from the data itself.
In particular an hypothetical systematic error of ~1\% on the core width of multiple scattering
can be easily fitted from data, with negligible impact on the fitted
parameters of interest and minor effect on the final accuracy.
This is clearly understood: a safe control region exists in phase
space where the systematics produce large distortions while the signal is vanishingly small.
Further studies are ongoing introducing many nuisance parameters corresponding
to the individual systematic effects to be fitted simultaneously.

\newpage 

\subsection[Analytic Evaluation of the NNLO virtual corrections to Muon--Electron scattering]{Analytic Evaluation of the NNLO virtual corrections to \\Muon--Electron scattering}
\addtocontents{toc}{\hspace{2cm}{\sl M.K.~Mandal}\par}
\vspace{5mm}

\noindent
{\normalsize 
{Manoj K. Mandal}
\footnote{Speaker, on behalf of R. Bonciani, A. Broggio, S. Di Vita, A. Ferroglia, P. Mastrolia, L. Mattiazzi, A. Primo, J. Ronca, U. Schubert, W.J. Torres Bobadilla, F. Tramontano; based on Ref.~\cite{Bonciani:2021okt}. }
,
{Luca Mattiazzi}
\\
{\footnotesize {\it INFN, Sezione di Padova, Via Marzolo 8, 35131 Padova, Italy}}
\\
{Jonathan Ronca}
\\
{\footnotesize \it Dip.to di Fisica, Universit\`a di Napoli Federico II and INFN, Sezione di Napoli, I-80126 Napoli, Italy}
\\
{William J.~Torres~Bobadilla}
\\{\footnotesize \it Max-Planck-Institut f{\"u}r Physik, Werner-Heisenberg-Institut, 80805 M{\"u}nchen, Germany}
}
%
%
%
\vspace{5mm}

\begin{figure}[t] 
    \centering
    \includegraphics[scale=0.45]{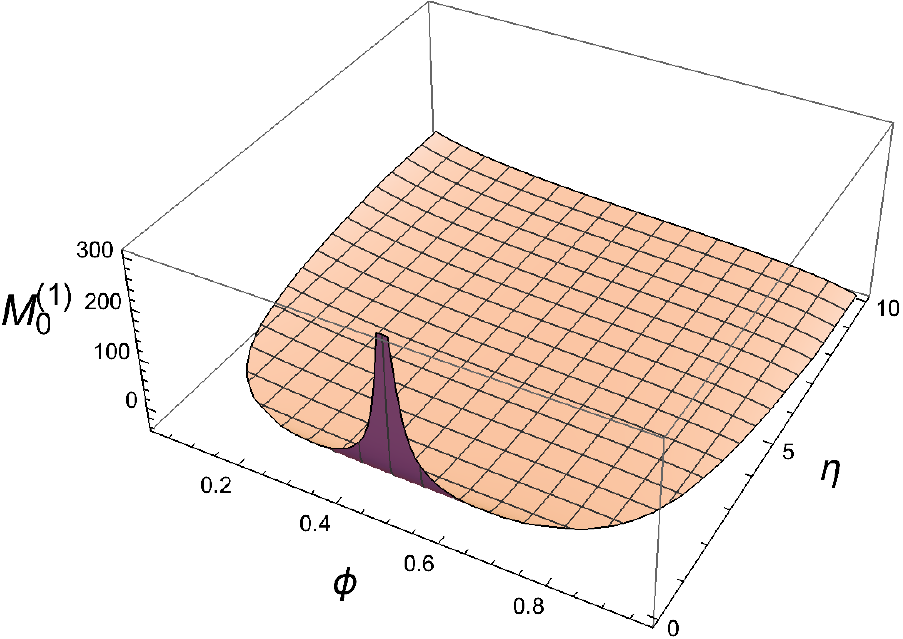}
    \includegraphics[scale=0.45]{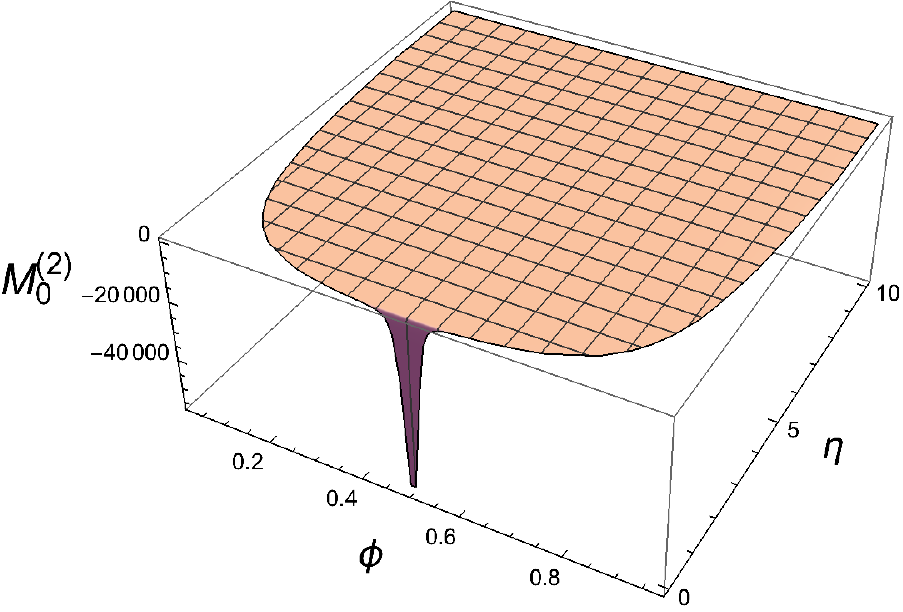}    	
	\caption{Three-dimensional plots of finite terms
		of the renormalized one- and two-loop amplitudes.
		Here, $\eta = s/(4M^2) - 1$, $\phi = -(t-M^2)/s$.
	}
	\label{fig:3d_plots_total1L2L}
\end{figure}

This contribution is based on the work carried out in Ref.~\cite{Bonciani:2021okt}. The recent results of the FNAL Muon $g-2$ experiment~\cite{Abi:2021gix} have confirmed the long standing discrepancy between the Standard Model (SM) prediction and the experimental measurement of the anomalous magnetic moment of the muon. The theoretical uncertainty in the SM prediction is dominated by the hadronic contribution.
Recently, a novel experiment, MUonE~\cite{Abbiendi:2016xup, Calame:2015fva}, has been proposed to determine this hadronic contribution using the elastic scattering of Electron and Muons.
%
This, in turn, demands the Muon--Electron scattering 
at Next-to-Next-to Leading order (NNLO) accuracy in the electromagnetic coupling~\cite{Banerjee:2020tdt}.
Here, we discuss the first step towards this direction taken in Ref.~\cite{Bonciani:2021okt}, where the complete analytic evaluation of the scattering amplitude of a pair of massless Electrons to a massive pair of Muons at the two-loop level in Quantum Electrodynamics (QED) was reported.
We start considering the production of a pair of Muons $e^{-}(p_1) + e^{+}(p_2) \rightarrow \mu^{-}(p_3) + \mu^{+}(p_4)$,
with $m_e=0$ and $m_{\mu} = M \ne 0$. Upon crossing, we obtain the Muon--Electron scattering amplitude.
The Mandelstam invariants are defined as
$	s=(p_1+p_2)^2, \, \, t=(p_1-p_3)^2, $ and $ \, u=(p_2-p_3)^2 $.
There are 6 diagrams at one loop and 69 diagrams at two-loop. 
We prepare the amplitude using an in-house software, 
built with an interface to FeynArts~\cite{Hahn:2000kx}, and
FeynCalc~\cite{Shtabovenko:2020gxv}. 
We perform the {\it adaptive integrand decomposition}~\cite{Mastrolia:2016dhn}, 
to simplify the integrands further. 
Then, we perform the Integration-By-Parts reduction via the interface to {\sc Reduze}~\cite{vonManteuffel:2012np}, and {\sc KIRA}~\cite{Maierhoefer:2017hyi} to obtain the amplitude in terms of independent integrals, known as the Master Integrals (MIs). All of these steps has been automatized in the {\sc Aida} framework~\cite{Mastrolia:2019aid}. 
The analytical experssion of MIs~\cite{Bonciani:2008az,Bonciani:2009nb,Mastrolia:2017pfy,DiVita:2018nnh,Becchetti:2019tjy} are known and they are expressed via Generalized Polylogarithms(GPLs). The two-loop amplitude contains 4063 GPLs up to weight four, whose arguments are written in terms of 18 different letters. After performing the Ultra-violet renormalization, the remaining Infrared (IR) poles of the one and two-loop renormalized amplitudes agree with the universal IR prediction~\cite{Becher:2009qa,Becher:2009kw}, independently obtained within the framework of {\it soft-collinear effective theory}. 
We show the finite part of the renormalized amplitudes in the physical region in Fig.~\ref{fig:3d_plots_total1L2L}.

These analytic results can be directly used to study the massive Muon pair production in massless electron annihilation, and to the study of the elastic scattering of Muon--electron scattering at NNLO accuracy in QED.

\clearpage 

\subsection[MC developments for $\mu e$ scattering at 10ppm]{MC developments for $\boldsymbol{\mu e}$ scattering at 10ppm}
\addtocontents{toc}{\hspace{2cm}{\sl T.~Engel}\par}

\vspace{4mm} Tim Engel for the {\sc McMule} Team

\vspace{4mm}

\noindent
Paul Scherrer Institut, CH-5232 Villigen PSI, Switzerland \\
Universit\"at Z\"urich, Winterthurerstrasse 190, CH-8057 Z\"urich, Switzerland
 \\
\vspace{4mm}

Precision experiments with leptons are an essential part of the search of physics beyond the Standard Model. In order to meet the experimental precision the calculation of next-to-leading order (NLO) or even next-to-next-to-leading order (NNLO) QED corrections has become mandatory. This has triggered the development of {\sc McMule}, a Monte Carlo for muons and other leptons. A special focus lies thereby on muon--electron scattering due to the MUonE experiment requiring a high-precision theory prediction at the level of 10ppm. In addition to that, the multi-scale nature of $\mu e$ scattering makes the corresponding calculation also interesting from a technical point of view. On the one hand, the regularisation of collinear divergences by finite fermion masses greatly simplifies the infrared structure compared to QCD. On the other hand, it also results in an increased complexity in the calculation of loop integrals. Furthermore, the small electron mass typically gives rise to strong scale hierarchies complicating a numerically stable implementation of amplitudes. The {\sc McMule} framework is based on the philosophy of exploiting the simplicities of QED while at the same time solving the corresponding ubiquitous problems in a process independent way.

Using the simple exponentiating structure of soft singularities in QED we have developed a subtraction scheme to all orders in perturbation theory~\cite{Engel:2019nfw}. This FKS$^{\ell}$ scheme allowed us to calculate the dominant electron line corrections for $\mu e$ scattering at NNLO~\cite{Banerjee:2020rww}. While perfect agreement was found with the calculation of Ref.~\cite{CarloniCalame:2020yoz}, significant discrepancies were observed when comparing to Ref.~\cite{Bucoveanu:2018soy}. For the full set of NNLO corrections we can rely on the recently computed two-loop amplitude with a non-vanishing muon mass~\cite{Bonciani:2021okt}. However, finite electron mass effects have to be taken into account as well in the calculation of physical observables. For this reason we have extended the method of `massification' to heavy external states~\cite{Engel:2018fsb}. Due to the universality of collinear degrees of freedom massification determines all mass effects that are not polynomially suppressed based on the massless amplitude. These developments leave a numerically stable implementation of the real-virtual contribution as the remaining bottleneck. In the case of Bhabha and M{\o}ller scattering~\cite{Banerjee:2021mty, Banerjee:2021qvi} we have found an elegant solution to this problem by expanding the amplitude in the soft photon momentum up to and including subleading power. We therefore expect this `next-to-soft stabilisation' to be immensely helpful for $\mu e$ scattering as well. Even more so since we have recently  systematised the corresponding calculation by extending the Low-Burnett-Kroll theorem to one loop~\cite{Engel:2021ccn}. This makes it trivial to determine the subleading term in the soft expansion from the non-radiative amplitude.

The FKS$^{\ell}$ subtraction scheme and the methods of massification and next-to-soft stabilisation are efficient solutions to the main difficulties related to the calculation of the full set of NNLO QED corrections to $\mu e$ scattering. The corresponding results will therefore be available in the near future, representing a milestone towards the 10ppm goal of the MUonE experiment.

\newpage 

\subsection{Towards muon--electron scattering at NNLO}
\addtocontents{toc}{\hspace{2cm}{\sl E.~Budassi}\par}

\vspace{5mm} Ettore Budassi

\vspace{5mm}

\noindent
University of Pavia - INFN Pavia
 \\
\vspace{5mm}

The anomalous magnetic moment of the muon $a_\mu=(g-2)/2$ is a fundamental quantity in particle physics. It has recently been measured at the Brookhaven
National Laboratory in 2001 \cite{bennett:2006fi} and at the Fermilab Muon $g-2$ Experiment
(E969) \cite{Abi:2021gix}. The results deviate by 4.2$\sigma$ from the Standard Model predictions \cite{Aoyama:2020ynm} where the most dominant source of theoretical error comes from the leading-order hadronic-vacuum-polarization contribution. More recently, an estimate of the LO HVP contribution has been proposed with Lattice QCD approach, apparently reducing the discrepancy with the measurements to about 1.5$\sigma$ \cite{Borsanyi:2020mff}.

Thus, a third and independent determination of the LO HVP contribution to the muon anomalous magnetic moment $a_\mu^\text{HVP, LO}$ is crucial to shed light on the discrepancy \cite{Calame:2015fva}. For these reasons, the MUonE experiment has been proposed at CERN. The experiment will exploit elastic muon--electron scattering data at relatively low momentum transfer \cite{Abbiendi:2016xup}. In order to match the precision of current theoretical predictions for $a_\mu^\text{HVP,LO}$, the required accuracy is of about 10 parts per million on the differential cross section. 

From the theoretical point of view, this implies the inclusion of all the relevant radiative corrections, as reviewed in Ref.~\cite{Banerjee:2020tdt}. Since recently, a Monte Carlo event generator, \textsc{mesmer}, is under active development to account for such contributions. 

Using this generator, all Next-to-Leading Order (NLO) contributions in the Electroweak theory have been numerically calculated \cite{Alacevich:2018vez}. The next step towards the 10 ppm goal on the differential cross section was to implement the fixed-order photonic Next-to-Next-to-Leading Order (NNLO) corrections. This has been done exactly for the virtual photon insertion on a single leg and for all the real photon emission. An approximate calculation, following the Yennie-Frautschi-Suura approximation, has been performed for the virtual photon insertion on the NLO boxes: all these corrections are at the \% level with respect to the tree level process \cite{CarloniCalame:2020yoz}. Another set of important corrections at this perturbative order are the NNLO Lepton Pair corrections. Such terms have been exactly calculated using the \textsc{mesmer} event generator and contribute in a range between $10^{-4}$ and some \% in some phase space regions \cite{Budassi:2021twh}.

Such calculations are crucial to achieve the precision goal of MUonE and need to be implemented into a fully exclusive Monte Carlo event generator for data analysis. However, it is crucial to tackle even higher order corrections in order to reach the goal of 10 ppm on the differential cross section.

\newpage 

\subsection[Hadronic contribution to the muon anomalous 
magnetic moment within DPT]{Hadronic contribution to the muon anomalous \\ 
magnetic moment within~DPT}
\addtocontents{toc}{\hspace{2cm}{\sl A.V.~Nesterenko}\par}

\vspace{5mm} A.V.~Nesterenko

\vspace{5mm}

\noindent
Joint Institute for Nuclear Research, Dubna, Russian Federation
\\
\vspace{5mm}

\begin{figure}[h]
    \centering
    \includegraphics[width=70mm,clip]{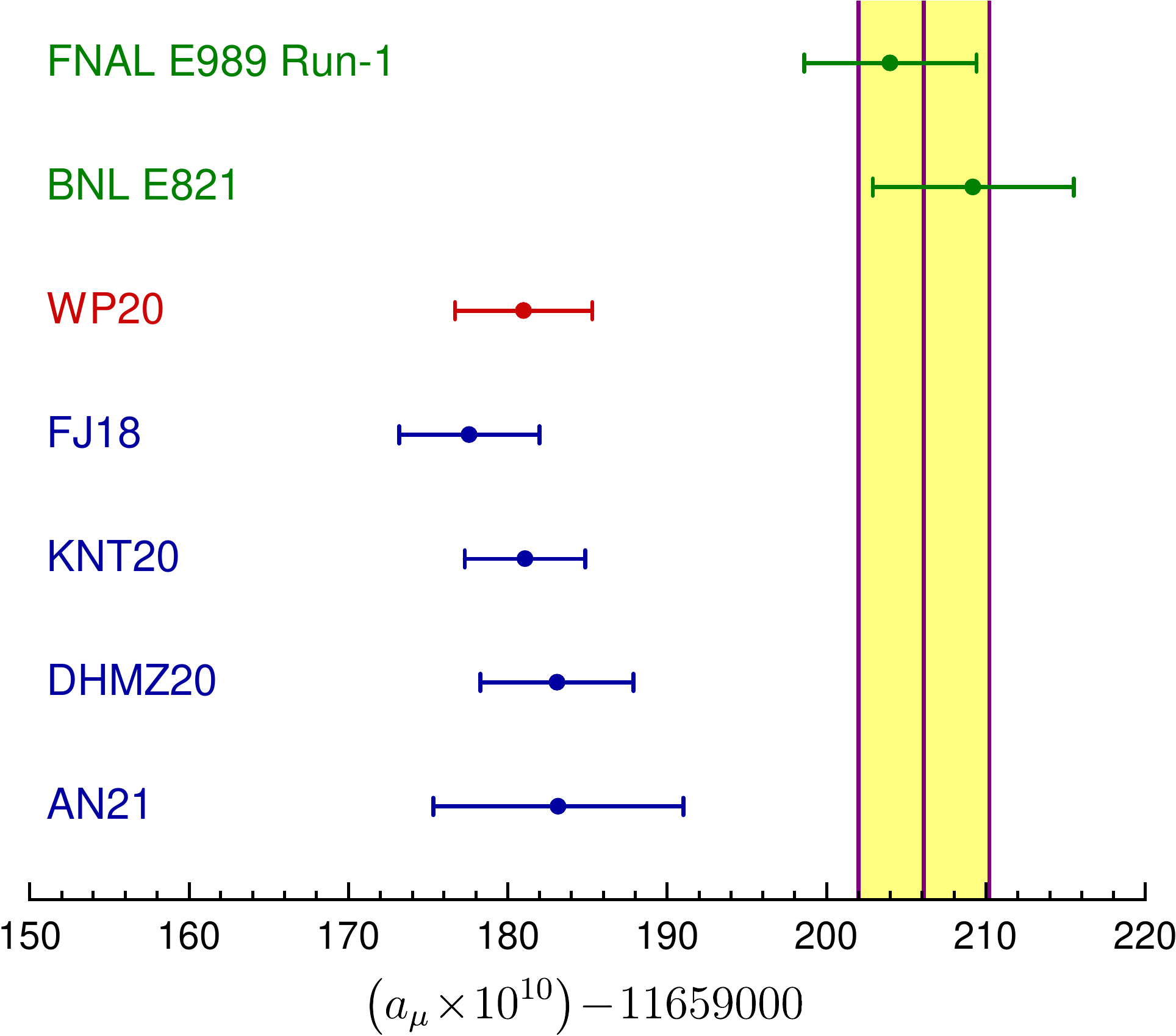}
    \hspace{12.5mm}
    \raisebox{5mm}{\includegraphics[height=57mm,clip]{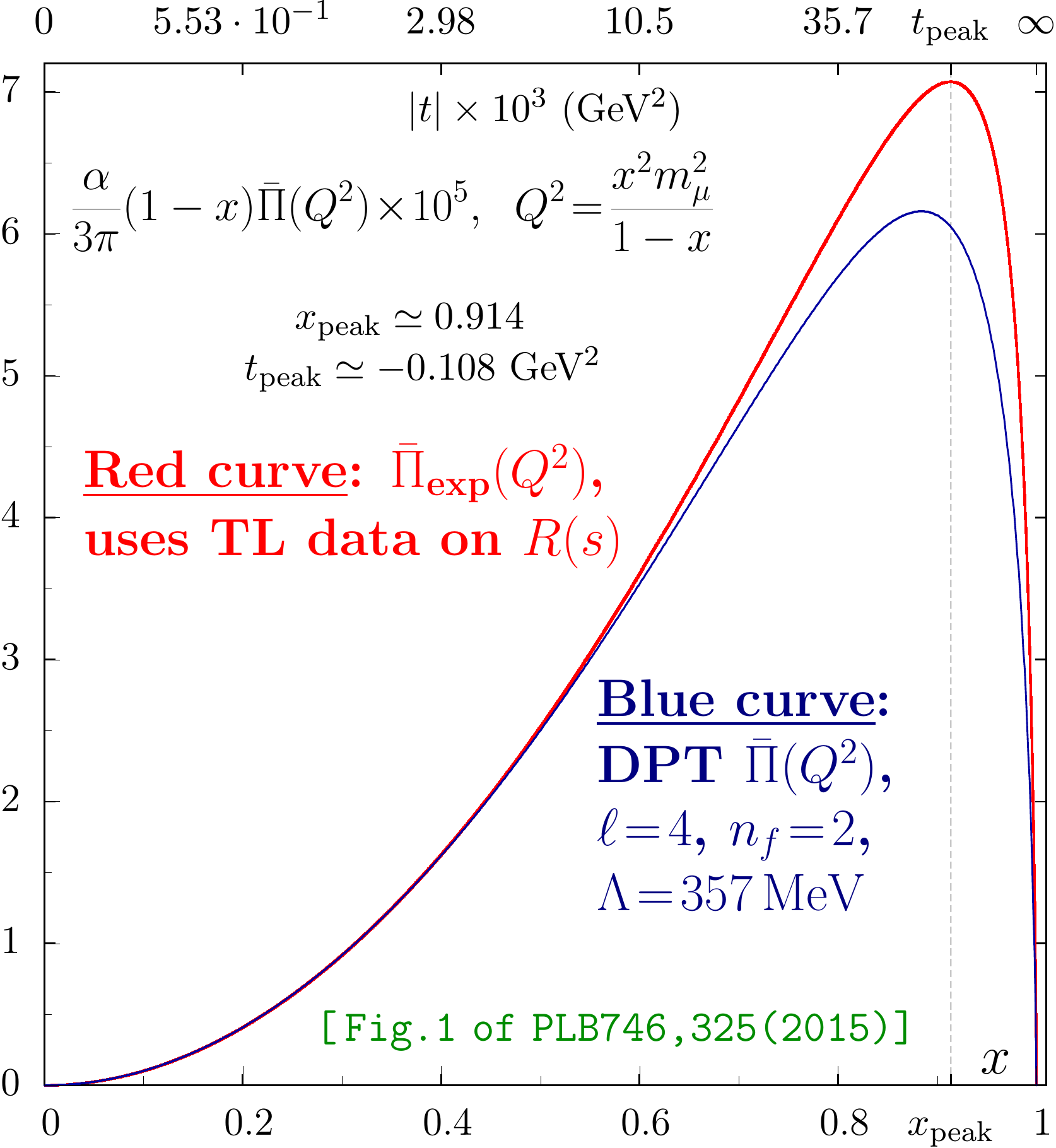}}
    \caption{}
    \label{fig:Nesterenko}
\end{figure}

Hadronic contribution to the muon anomalous magnetic moment is studied 
within dispersively improved perturbation theory (DPT)~\cite{Nesterenko:2013vja,Nesterenko:2014txa,Nesterenko:2016pmx}. The latter combines in a self--consistent way the QCD perturbation 
theory with the nonperturbative constraints originated in the corresponding 
dispersion relations. The obtained hadronic vacuum polarization function 
$\bar\Pi(Q^2)$ eventually yields~$a_{\mu}^{\text{HVP, LO}} = 
(695.1 \pm 7.6) \times 10^{-10}$~\cite{AmuDPT} (with the world average 
value of~$\alpha_{s}(M_{Z}^{2})$~\cite{ParticleDataGroup:2020ssz} being used), which is in 
a good agreement with recent estimations of the quantity on~hand. The 
corresponding Standard Model prediction [$a_{\mu} = (11659183.2 \pm 7.8) 
\times 10^{-10}$] is~in the right ballpark and differs from the combined 
BNL--FNAL experimental measurement by~$2.6$~standard deviations (Fig.~\ref{fig:Nesterenko}, left). Additionally, the DPT hadronic vacuum polarization function 
$\bar\Pi(Q^2)$ can be employed as a supplementing infrared input for the 
MUonE project~\cite{Abbiendi:2016xup,Calame:2015fva} (particularly, in the energy range uncovered 
by measurements) and the lattice studies (Fig.~\ref{fig:Nesterenko}, right), see 
Ref.~\cite{AmuDPT} for the details.

\newpage 

\subsection{Preparing for MUonE experiment---what can we learn from lattice and dispersive data?}
\addtocontents{toc}{\hspace{2cm}{\sl M.K.~Marinkovi{\'c}}\par}

\vspace{5mm} Javad Komijani, Marina Krsti{\'c} Marinkovi{\'c}\footnote{Speaker}{~}

\vspace{5mm}

\noindent
Institut f\"ur Theoretische Physik, ETH Zurich, 
Wolfgang-Pauli-Str. 27, 8093 Z\"urich, Switzerland
 \\

\vspace{5mm}

Recent results by the Muon $g-2$ experiment at Fermilab~\cite{Abi:2021gix}, when
combined with the results by the BNL's E821
experiment~\cite{bennett:2006fi},
solidified the long{-}standing discrepancy between the experimental
observations and theoretical prediction of the muon $g-2$, {wherein the leading hadronic contributions
come from a}  data{-}driven {calculation that is based
on a dispersive} 
approach. 
{To improve the theoretical prediction corresponding to hadronic vacuum
polarization (HVP),}
{p}arallel to the lattice communities effort in obtaining results
comparable to the precision of the recent subpercent lattice {calculation}~\cite{Borsanyi:2020mff}, 
{an} alternative space-like {data-driven}
approach {is} proposed~\cite{Calame:2015fva,Abbiendi:2016xup};
the experimental exploration {within MUonE experiment} is underway~\cite{MUonE:LoI}. The experimental strategy and theoretical foundations of the MUonE experiment {are} discussed in several talks  at this workshop.
We recall here the advantage of the space-like over the time-like method that requires meticulous accounting of numerous exclusive hadronic channels, 
while the former features a smooth HVP integrand~\cite{Lautrup:1971jf,Blum:2002ii}. 

Proposed template fit strategy~\cite{MUonE:LoI} relies on the parametrization inspired by the contribution of lepton-pairs and top quarks to the  space-like photon vacuum polarization
\begin{equation}
\Delta{\alpha_{\mathrm{had}}(t)}=k\Bigg\{ -\frac{5}{9} - \frac{4M}{3t} 
+ \Big ( 
\frac{4M^2}{3t^2} + \frac{M}{3t} -\frac{1}{6}
\Big) 
\frac{2}{\sqrt{1-\frac{4M}{t}}} \log 
\Bigg| \frac{1-{\sqrt{1-\frac{4M}{t}}}}{ 1+{\sqrt{1-\frac{4M}{t}}} } \Bigg |
\Bigg\}.
\label{eq:lepton}
\end{equation}
A natural alternative to this {particular} parametrization would include  Pad{\'e} approximants (PA). {W}riting the HVP in terms of a Stieltjes function warrants an existence of the converging sequence of {order} $[N-1,N]$ and  $[N,N]$ PAs~\cite{Aubin:2012me}, defined as 
\begin{eqnarray}
 \Delta{\alpha_{\mathrm{had}}(Q^2)} = 
c_0+ Q^2 \Bigg(a_0+\sum_{i=1}^{N}  \frac{a_i}{b_i+Q^2}\Bigg),
\label{eq:Pade}
\end{eqnarray}
where $Q^2=-t$, $a_0=0$ in $[N-1,N]$  PAs and is left as a fitting parameter in  $[N,N]$ PAs. 

The dispersive $\tau-$based  isospin-one  model of the hadronic vacuum polarization has been instrumental in understanding the systematics of the  Pad{\'e} fits in early lattice QCD calculations~\cite{Golterman:2014ksa}, among other results rendering 
choices for fit functions 
to the lattice data. (Another choice is vector meson dominance, which is now
{obsolete because they lead to}
large systematic errors). 
We propose to use this constrained model of the $I=1$ vector-channel polarization function to understand potential effect of the chosen fit function to the MUonE experiment. Prior to the actual MUonE data 
{being} available, we resort to the ALEPH experiment data\footnote{$I=1$ HVP data from ALEPH experiment
were kindly provided by Kim Maltman and collaborators~\cite{Golterman:2014ksa}.}. 
Our preliminary results indicate that the lepton-like ansatz gives a result for the $I=1$ HVP in the momenta range $[0,1]\mathrm{GeV}^2$ at least as good as the higher order  Pad{\'e}s conventionally favoured by the lattice data, when compared with the value obtained by the numerical integration. 
The understanding of the reason of such good performance of the lepton-like ansatz in Eq.~\eqref{eq:lepton} necessitates further investigation. 

\newpage

\subsection[Lattice calculation of the short and intermediate time windows contributing to the leading-order HVP term of the muon $g-2$ using twisted mass fermions]{Lattice calculation of the short and intermediate time windows contributing to the leading-order HVP term of the muon $\boldsymbol{g-2}$ using twisted mass fermions}
\addtocontents{toc}{\hspace{2cm}{\sl G.~Gagliardi}\par}

\vspace{5mm} Giuseppe Gagliardi (On behalf of the ETM Collaboration)

\vspace{5mm}

\noindent
Istituto Nazionale di Fisica Nucleare, Sezione di Roma Tre,
Via della Vasca Navale 84, I-00146 Rome, Italy.
 \\

\vspace{5mm}

\begin{figure}[h]
  \includegraphics[width=0.49\linewidth]{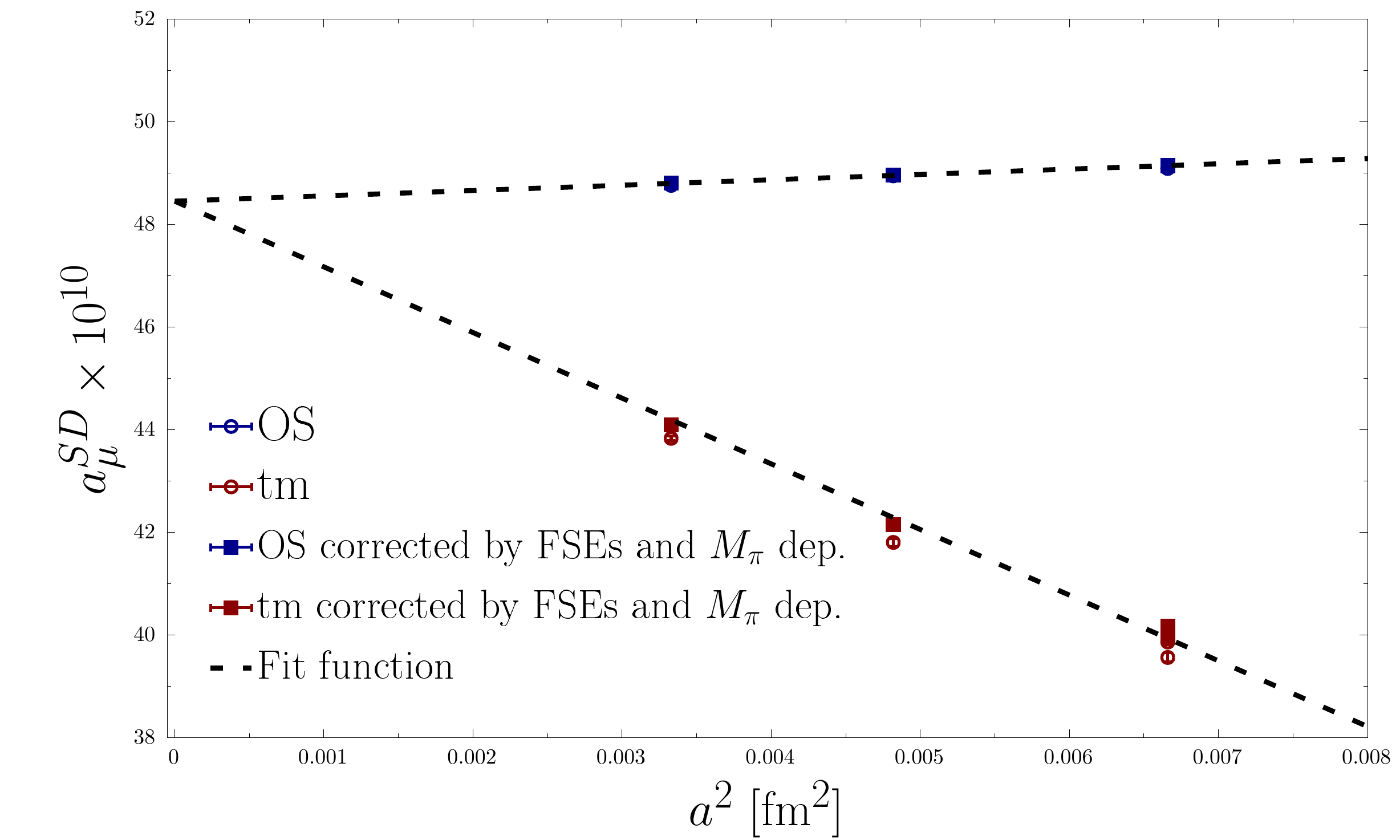}  
  \includegraphics[width=0.49\linewidth]{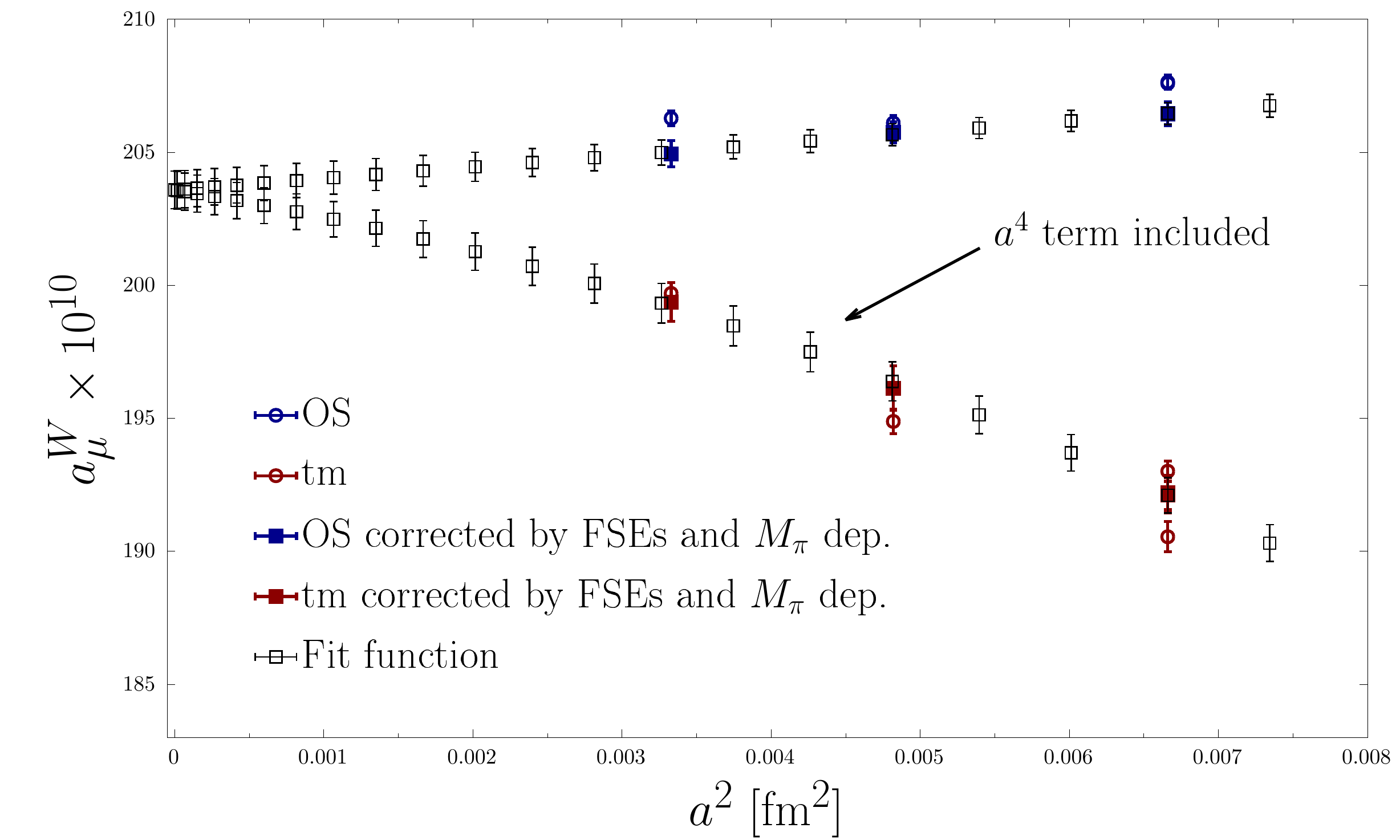}  
  \caption{}
  \label{fig:ETM}
\end{figure}

We present our preliminary results for the light-connected contribution to the short distance ($a_{\mu}^\text{SD}$) and intermediate time ($a_{\mu}^\text{W}$) windows contributing to $a_{\mu}^\text{HVP}$~\cite{RBC:2018dos}. The calculation has been performed using some of the pure QCD gauge ensembles produced by the ETMC with $N_{f}=2+1+1$ Wilson-clover twisted mass fermions~\cite{ExtendedTwistedMass:2021qui,ExtendedTwistedMass:2021gbo}. The four ensembles we considered correspond to almost physical pion masses, and lattice spacing from $0.082~{\rm fm}$ down to $0.058~{\rm fm}$. Two ensembles differing only on the lattice size ($L\sim 5.2~{\rm fm}, L\sim7.8~{\rm fm}$) have been used to control Finite Size Effects (FSEs). We made use of two versions of the local vector current $J_{\mu}$, namely\\ \vspace{-1.3cm}
\begin{center}
\begin{align}
J_{\mu}^{OS}(x) \propto ~Z_{V}~\bar{\psi}_{\ell}^{+}(x)~\gamma_{\mu}~\psi_{\ell}^{+}(x),\qquad  J_{\mu}^{tm}(x) \propto ~Z_{A}~\bar{\psi}_{\ell}^{+}(x)~\gamma_{\mu}~\psi_{\ell}^{-}(x)~,\nonumber~
\end{align}
\end{center}
which correspond to equal (OS) or opposite (tm) values of the Wilson $r-$parameters of the two quarks.
The results obtained using the two currents differ by $O(a^{2})$ lattice artifacts, allowing to approach the continuum limit in two different ways. The corresponding renormalization constants have been determined with sub-permille precision making use of Ward identities.
In the two windows, the contribution from $\langle J_{i}(t)J_{i}(0)\rangle$ at large time distances $t$, which is affected by sizable FSEs and by a signal-to-noise problem, is exponentially suppressed, and the main issue is to perform a reliable continuum limit extrapolation. This is particularly important for $a_{\mu}^\text{SD}$, where dangerous $a^{2}\log{(1/a)}$ lattice artifacts are already generated in the free theory~\cite{Harris:2021azd}. These  artifacts, which slow down the convergence, have been subtracted from our lattice data, after an explicit calculation in lattice perturbation theory at leading order. We performed several fits in order to correct for FSEs, mistuning of the pion masses and to extrapolate to the continuum limit. Our preliminary results for $a_{\mu}^\text{SD}$ and $a_{\mu}^\text{W}$ are (see Fig.~\ref{fig:ETM})
\begin{align}
a_{\mu}^\text{SD} = 48.35(24)\times 10^{-10},\qquad a_{\mu}^\text{W} = 204.2(1.2)\times 10^{-10}~,  \nonumber
\end{align}
where the errors are inclusive of a preliminary estimate of the systematics.

\clearpage

\section{Program committee}
\begin{itemize}
    \item Achim Denig - JGU Mainz

\item Fedor Ignatov - BINP Novosibirsk 

\item Andrzej Kupsc - Uppsala University and NCBJ Warsaw

\item Alberto Lusiani - SNS Pisa

\item Bogdan Malaescu - LPNHE, CNRS Paris

\item Stefan Mueller - HZDR Dresden

\item Massimo Passera - INFN Padova

\item Thomas Teubner - University of Liverpool

\item Graziano Venanzoni - INFN Pisa
\end{itemize}

\section{List of participants}
See \url{https://agenda.infn.it/event/28089/registrations/participants}

\section{Acknowledgements}
This work was supported by the European Union STRONG 2020 project under Grant Agreement Number 824093, Fellini
Fellowship for Innovation at INFN funded by the European Union’s Horizon 2020 research and innovation programme under the Marie
Skłodowska-Curie grant agreement No 754496, the STFC Consolidated Grant ST/T000988/1, and the SNSF (Project No.\ PCEFP2\_181117 and contract 200021\_178967).

\newpage

\bibliographystyle{apsrev4-1_mod} 
\bibliography{main}

\end{document}